\documentclass[12pt, preprint]{aastex}
\usepackage{afterpage}
\usepackage{graphicx}
\shorttitle{Band-limited Imaging}
\shortauthors{Fruchter}
\begin{document}

%-----------------------------------------------------------------------
%		            Paper Title 
%-----------------------------------------------------------------------
% Enter the title of the paper.
%
% EXAMPLE: \title{A Breakthrough in Astronomical Software Development}
%
% If your title is so long as to fill the page header when you print it,
% then please supply a short form as a \titlemark.  You don't need a
% \titlemark macro if your title fits in the page header.
%
% EXAMPLE:
%  \title{Rapid Development for Distributed Computing, with Implications
%         for the Virtual Observatory}
%  \titlemark{Rapid Development for Distributed Computing}

\title{A New Method for Band-limited Imaging with Undersampled Detectors}
%\titlemark{}

%-----------------------------------------------------------------------
%		          Authors of Paper
%-----------------------------------------------------------------------
% Enter the authors followed by their affiliations.  The \author and
% \affil commands may appear multiple times as necessary.  List each
% author by giving the first name or initials first followed by the
% last name.  Authors with the same affiliations should grouped
% together. 
%
% Try to limit the front matter to no more than three \author
% commands.  Group authors with the same affiliations.  Too many
% \author commands fills the first page of the paper with little
% actual text.
%
% EXAMPLE:
% \author{S. Djorgovski\altaffilmark{1,2} and Ivan R. King}
% \affil{Astronomy Department, University of California,
%     Berkeley, CA 94720}
%
% \author{C. D. Biemesderfer\altaffilmark{3}}
% \affil{National Optical Astronomy Observatories, Tucson, AZ 85719}

\author{Andrew S. Fruchter}
\affil{Space Telescope Science Institute, Baltimore, MD 21218}

\hyphenation{Feich-tinger}
\begin{abstract}
Since its original use on the Hubble Deep Field, ``Drizzle" has become a de facto standard for the combination of  images taken by the Hubble Space Telescope.   However, the drizzle algorithm was developed with small, faint, partially-resolved sources in mind, and is not the best possible algorithm for high signal-to-noise unresolved objects.   Here, a new method for creating band-limited images from undersampled data is presented.   The method uses a drizzled image as a first order approximation and then rapidly converges toward a band-limited image which fits the data given the statistical weighting provided by the drizzled image.  The method, named iDrizzle, for
iterative Drizzle, effectively eliminates both the small high-frequency artifacts and convolution with an interpolant kernel  that can be introduced by drizzling.  The method works well in the presence of geometric distortion, and can easily handle cosmic rays, bad pixels, or other missing data.    It can combine images taken with random dithers, though the number of dithers required to obtain a good final image depends in part on the quality of the dither placements.  iDrizzle may prove most beneficial for producing high-fidelity point spread functions from undersampled images, and could be particularly valuable for future Dark Energy missions such as WFIRST and EUCLID, which  will likely attempt to do high precision supernova photometry and lensing experiments with undersampled detectors.  

\end{abstract}

%-----------------------------------------------------------------------
%			Subject Index keywords
%-----------------------------------------------------------------------
% Enter up to 6 keywords describing your paper.  These will NOT be
% printed as part of your paper; however, they will be used to
% generate the subject index for the proceedings.  There is no
% standard list; however, you can consult the indices for past Calibration
% Workshop Proceedings. 


%-----------------------------------------------------------------------
%			      Main Body
%-----------------------------------------------------------------------
% Place the text for the main body of the paper here.  You should use
% the \section command to label the various sections; use of
% \subsection is optional.  Significant words in section titles should
% be capitalized.  Sections and subsections will be numbered
% automatically. 
\section{Introduction}

Astronomical detectors are often undersampled.   A wide field-of-view is often most economically 
obtained by using large pixels, and large pixels produce a minimum of read noise in 
comparison to sky noise.     Large pixels, however, result in undersampling.   In order to fully sample an image with an undersampled detector one must dither and combine multiple images.   However, distortions in the field-of-view may make it impossible to perform shifts that equally well sample different parts of the detector.   In practice then the combined pixels from dithering of astronomical detectors often produce irregular sampling of the image plane.

\begin{figure}
\begin{center}
\includegraphics[width=3.0in]{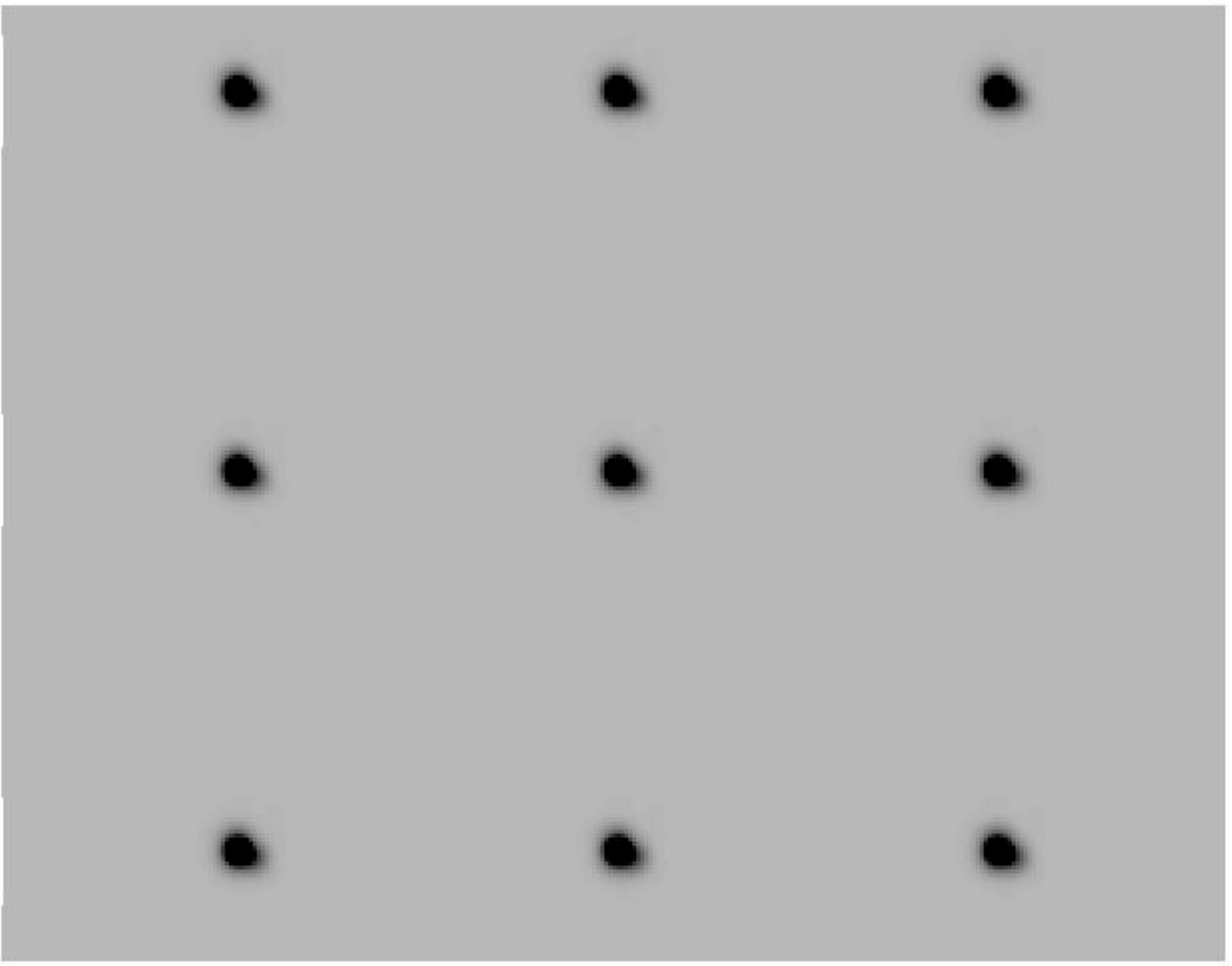}
\hspace{0.2in}
\includegraphics[width=3.0in]{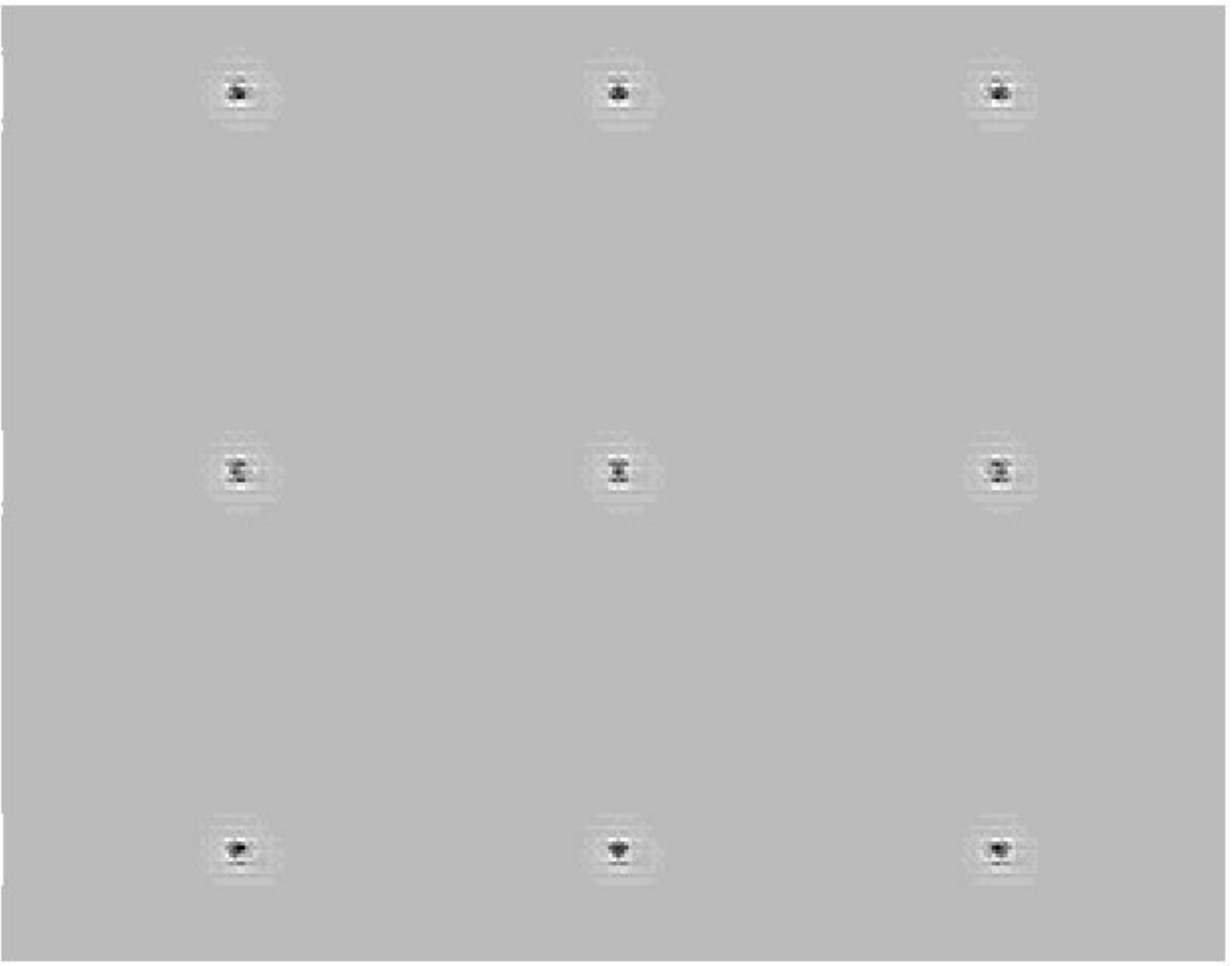}
%\plottwo{f1.eps}{f2.eps}
\end{center}
\caption{On the left a series of synthetic point spread functions (PSFs) for the Hubble Space Telescope (HST) Advanced Camera for Surveys (ACS).   On the right, a drizzled approximation of the image subtracted from the original.  ACS is relatively well sampled.  It has an optical PSF with full width at half-maximum (FWHM) of $\sim 0\farcs06$ in the R band with pixels of width $0\farcs05$.  Nonetheless, the largest residuals are still greater than 10\% of peak.  Drizzle smooths the PSF and adds high frequency noise.  The smoothing is reproducible; thus if a PSF slightly larger than the original is acceptable, the smoothing is not a significant issue.   Similarly, if one is measuring properties of the image on scales larger than a couple of original pixels, the high frequency noise largely averages out.   Thus Drizzle is well suited for aperture photometry with aperture radii larger than $\sim 2$ original pixels, as well as galaxy photometry.   If one wishes to reconstruct a true instrumental PSF, or do PSF fitting, another method would be preferred.
}
\label{fig-ACS-driz}
\end{figure}

In order to combine the irregularly sampled data from the Hubble Deep Field {\it HDF} (Williams et al. 1996), a new image
algorithm, Drizzle (Fruchter and Hook 2002) was developed.   Drizzle combines dithered images in a statistically optimal fashion.   However, as can be seen in Figure~\ref{fig-ACS-driz}, Drizzle adds small high-frequency artifacts to the image.   On scales larger than an original pixel, these rapidly average out.   Thus for the prime purpose of the {\it HDF}, the study of small faint galaxies, Drizzle is an excellent algorithm.  However, for the analysis of high signal-to-noise images of point sources, or other cases where preservation of the true point spread function (PSF) is essential, one might prefer an algorithm which more exactly reproduces the highest frequency features in the image.   

In order to produce higher-fidelity images than are created by Drizzle, Lauer (1999) introduced a method which attempts to analytically predict the values of a regularly subsampled image based on the values of dithered images.   This method works well when the dithers are nearly interlaced (that is when the dither positions nearly fall on an evenly spaced grid), or when the data can be broken up into multiple sets of nearly-interlaced datasets, though a loss in accuracy is seen as the samples diverge from interlacing.  It cannot handle substantial image distortions, however, but must instead create local reconstructions.   This reflects a general problem in the field -- there is no known analytical method for reconstructing a band-limited function from irregularly sampled data, even when that sampling everywhere reaches or exceeds the Nyquist rate.

\begin{figure}
\includegraphics[width=6.5in]{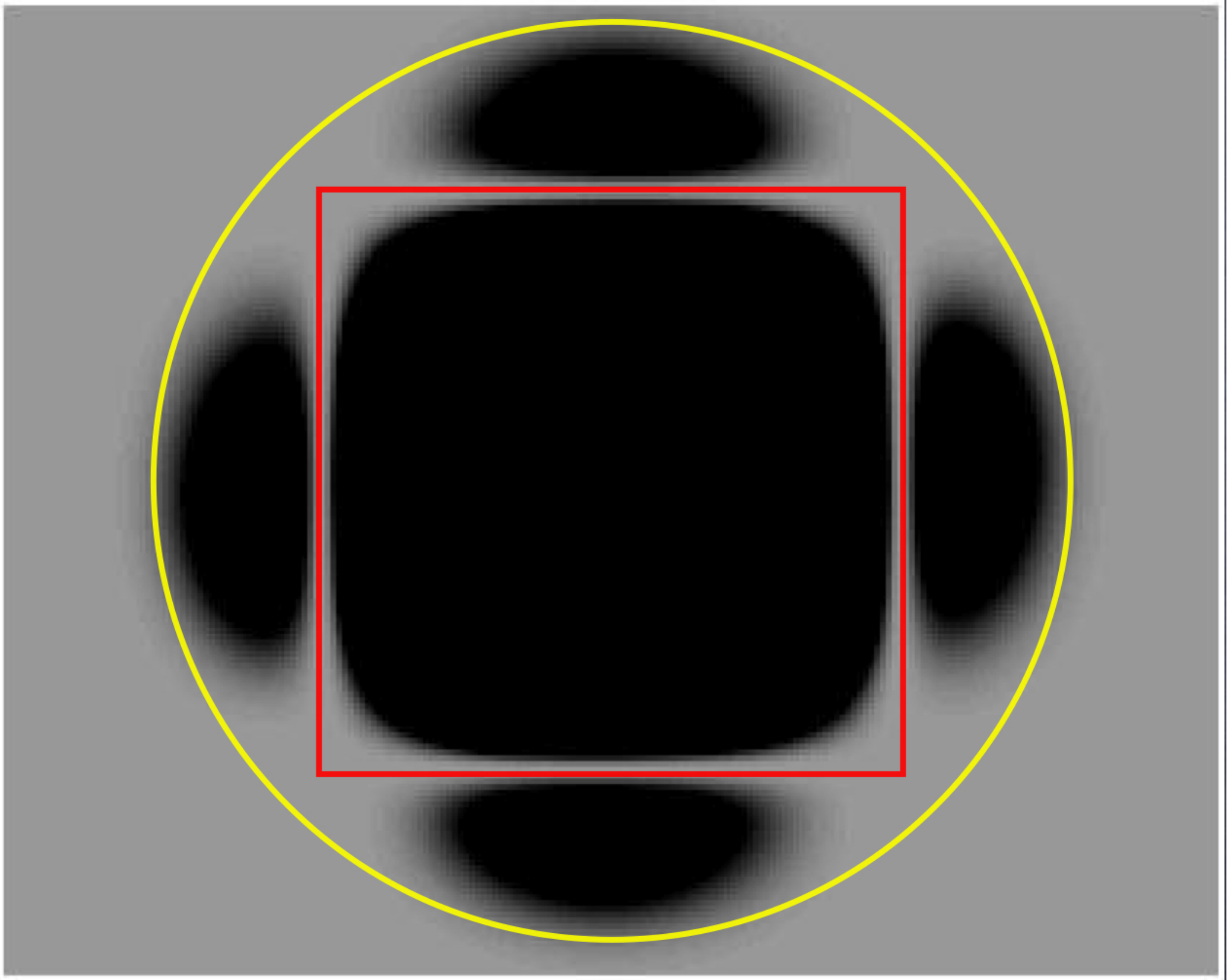}
%\plotone{f3.eps}
\caption{The power spectrum of the PSF wide-field camera of the HST Wide-Field and Planetary Camera 2 (WFPC2).   This imager has pixels $0\farcs1$ in width, while the FWHM of the native optical PSF is about $\sim 0\farcs06$ in the R band.  In this image, the yellow circle shows the optical band limit of the telescope, $\lambda / D$.  The red square shows the sinc suppression of  the power spectrum caused by the WFPC2 pixel.  A two-times oversampled WFPC2 image would Nyquist sample all the power contained in the box.   However, even with this well-placed four point dither, the power beyond the red box will remain aliased.  A number of imagers being discussed for future space-based wide-field
imaging are as undersampled as the WFPC2.}
\label{fig-wfpc2-psf}
\end{figure}

The spatial frequencies in an astronomical image are strongly limited by the optics of the telescope and convolution with the detector pixel.   The effective angular power frequency cutoff is typically no larger than $| \vec{k} | \la D/ \lambda_s$, where $D$ is the (maximum) aperture of the telescope and $\lambda_s$ is the shortest wavelength in the passband.   Optical images are therefore band-limited (see Figure~\ref{fig-wfpc2-psf}). Given a set of undersampled individual exposures which are distorted, and whose dithers therefore result in varying patterns of sampling across the field, solving for a single image consistent with all the exposures is equivalent to reconstructing a band-limited function from irregularly sampled data.  

The reconstruction of a band-limited function from regularly sampled data through sinc interpolation is a well know result of information theory due to Shannon (1948), though the result goes back to earlier work by Whittaker (1935) and in the Russian literature to Kotelnovikov (1933).
In the case of irregular sampling, obtaining an analytical representation of the function is more complicated and involves the use of a type of separable Hilbert space called a ``frame''.   An introduction to the use of Frame Theory to solve the irregular sampling problem can be found in Benedetto (1992).   

However, the Frame Theory solution for irregular sampling often leads to functional inversions that are numerically difficult and which, when the data are not well-oversampled, can be ill-conditioned, even in the absence of noise.   This produces very large computational costs (Werther 1999, Aldroubi and Gr\"ochenig 2001).   Thus the Frame Theory approach is rarely used in practice.  Instead, iterative reconstruction techniques have been developed to solve for a band-limited function in the case of irregularly sampled data (c.f.~Feichtinger and Gr\"ochenig 1994, Werther 1999, Gr\"ochenig and Strohmer 2001 ).   Here I present a new iterative reconstruction method.   In essence it is an improvement upon a simple but powerful iterative technique, the Voronoi or nearest neighbor approximation (c.f.~Werther 2004).   In the method introduced here,  the astronomical imaging algorithm, Drizzle, replaces the nearest neighbor approximation.  Drizzle allows this method to handle to geometric distortion, and combines the data using
the full statistical power of the individual images.   In the absence of noise this method converges directly to the band-limited image.  In the presence of noise, the method introduces a small increase in statistical noise, but the systematic high-frequency noise introduced by Drizzle (see Figure~\ref{fig-ACS-driz}) is removed.

\section{The Method}

 One might imagine that one could take the total set of irregularly sampled data which everywhere meets the Nyquist criterion, perform a direct Fourier transform, remove any frequencies above the cutoff frequency, and the use the inverse Fourier transform to arrive at the true band-limited image.  Unfortunately, a direct Fourier transform of irregularly-sampled data throws a great deal of power out of the original passband.  This simple method thus fails terribly.   It is in fact more effective to first put the data onto a regular Nyquist grid by simply taking the value of the nearest neighbor before doing the Fourier transform.   The inverted, band-limited function turns out to be a much truer approximation than the direct transform case.    This approximation is know as the Voronoi approximation.  
 
The Voronoi approximation is band-limited function and thus can be sinc interpolated to the irregular grid of the data.  One can therefore subtract the Voronoi approximation from the original function at all of the data points.  Furthermore, this smaller difference function is itself a band-limited function, so one can repeat the process and get a further refined approximation to the underlying band-limited function.  This procedure is known to converge geometrically (see Theorem 8.13 in Feichtinger and Gr\"ochenig 1994).   

The nearest neighbor approximation is, however, far from ideal for astronomical imaging.   Only one data sample is used at any point on the regular grid, even if several nearby data samples could provide information, and there is no means to weight the data according to its statistical significance.  Therefore in the method proposed here the nearest neighbor approximation is replaced by Drizzle in each iteration.    While Drizzle introduces small artifacts (as does the nearest neighbor method) the iterative comparison with the original data serves to remove these.

The proposed procedure is described in detail immediately below.    If the reader is not already familiar with the ideas discussed in
the previous paragraphs, and in particular the iterative use of the Voronoi approximation to obtain a band-limited image, following the flow of this procedure may not be easy.   To these readers I strongly recommend the truly excellent and fairly short web tutorial, ``A First Guided Tour on the Irregular Sampling Problem", by Tobias Werther (2004). After understanding this tutorial, it will be clear the procedure presented here is but a variation on a theme.

\begin{enumerate}

\item  Drizzle the $N$ dithered images of a field, $\{I^1, I^2, I^3... I^N\}$, onto an {\it oversampled} output grid, to produce the image $D_1$.   The $1$ in the subscript of $D$ represents that this is the first iteration.   The drizzling is performed using weight files which provide the statistical weights of the individual image and zero-out bad pixels, cosmic rays, and other significant defects.

\item Fourier transform the image $D_1$ to create $\tilde{D_1} = F(D_1)$, where $F()$ is the Fourier operator.

\item Produce a band-limited version of the transform of the image, $\tilde{B_1} = L * \tilde{D_1}$, where $L(\vec{k}) = 1$ when $\vec{k} = 0$, $L(\vec{k}) = 0$, when $\vec{k}$ lies outside of the band-limited region, and  $L(\vec{k})$ tapers to $0$ as  $\vec{k}$ approaches the boundary of the band-limited region.

\item Transform the Fourier plane back to the image plane to obtain the first approximation to the true image, $A_1 = F^{-1}(\tilde{B_1})$.

\item Map the first approximation to the combined image, $A_1$, back to the frames of the original individual images using {\tt blot} (Fruchter and Hook 2002), producing a series of approximations to the original images, $A^m_1 = blot(A_1 \Rightarrow I^m)$.

\item Subtract the blotted approximations from the corresponding original images to produce a series of new images $I^m_1 = I^m - A^m_1$

\item Return to the first step and now drizzle the new set of images $\{I^1_1, I^2_1, I^3_2... I^N_1\}$ to produce the image $D_2$.

\item Continue as before with one modification.  Now at Step 4, in the Nth iteration, $A_N = F^{-1}(\tilde{B_N}) + \sum_{j=1}^{N-1}A_j$, since for  $j>1$, $D_j$ does not estimate the ``true'' image, but rather the difference between the true image and previous approximation.  

\item After the iterations are complete, the final approximation is an oversampled image.   This can be sinc interpolated back down to critical sampling, or indeed to an even coarser scale if preferred.

\end{enumerate}

With each iteration, the algorithm creates a band-limited approximation to the true underlying function, and subtracts the approximation from the original data, leaving a difference function that must be approximated in the next iteration.   However, the difference function, like the original true function, is band-limited.  If the data points were on a regularly sampled grid, these iterations would not be necessary.   One could compute the function on a set of desired output points simply by using the sinc interpolation.   However, it is the fact the samples are irregular that makes this procedure necessary.    Although the method creates a band-limited function at each step, the need to approximate the values on a regular grid causes an error.   However, this error drops as the amplitude of the remaining function falls with each successive iteration.

\begin{figure}[!h]
\begin{center}
\includegraphics[width=3.0in]{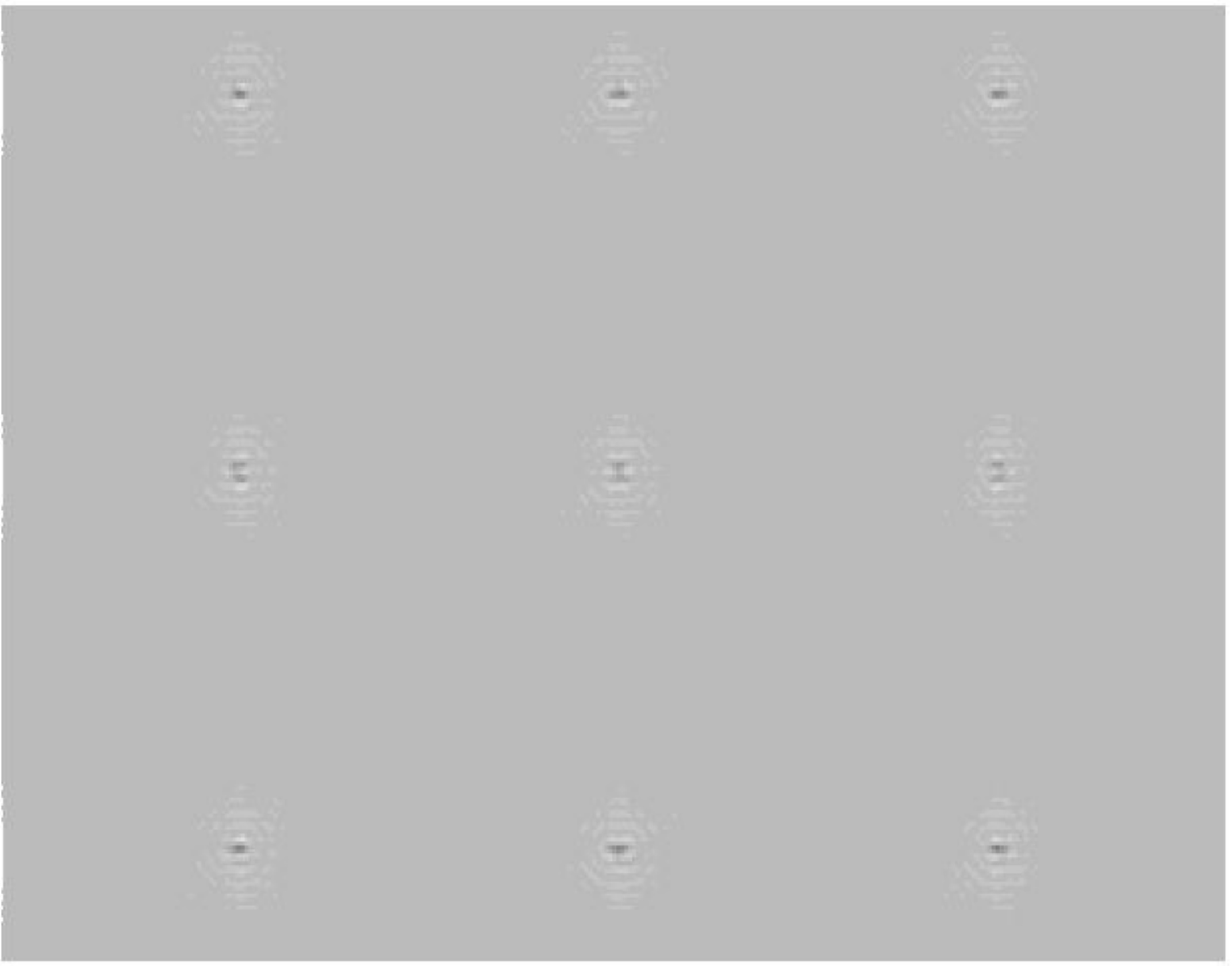}
\hspace{0.2 in}
\includegraphics[width=3.0in]{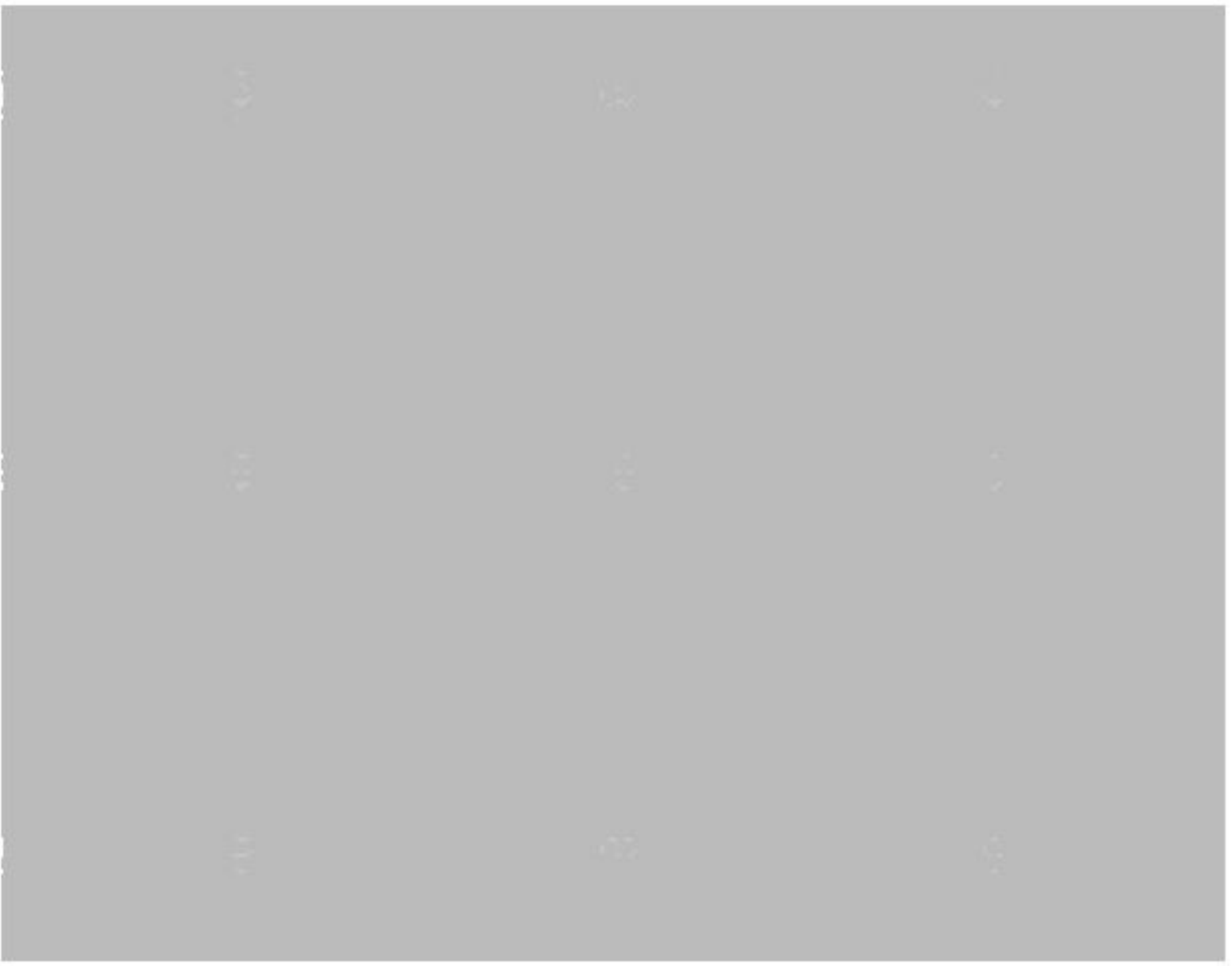}
\end{center}
%\plottwo{f4.eps}{f5.eps}
\caption{On the left, the residual difference between the true ACS PSF (see Figure~1) and the iterative approximation after the first iteration.   On the right the same difference after several more iterations.   The rapid reduction of the small artifacts of the Drizzle algorithm is evident.   Twelve simulated noiseless ACS images with random pointings were combined.}
\label{fig-ACS-iter}
\end{figure}

\begin{figure}
\begin{center}
\includegraphics[width=3.15in]{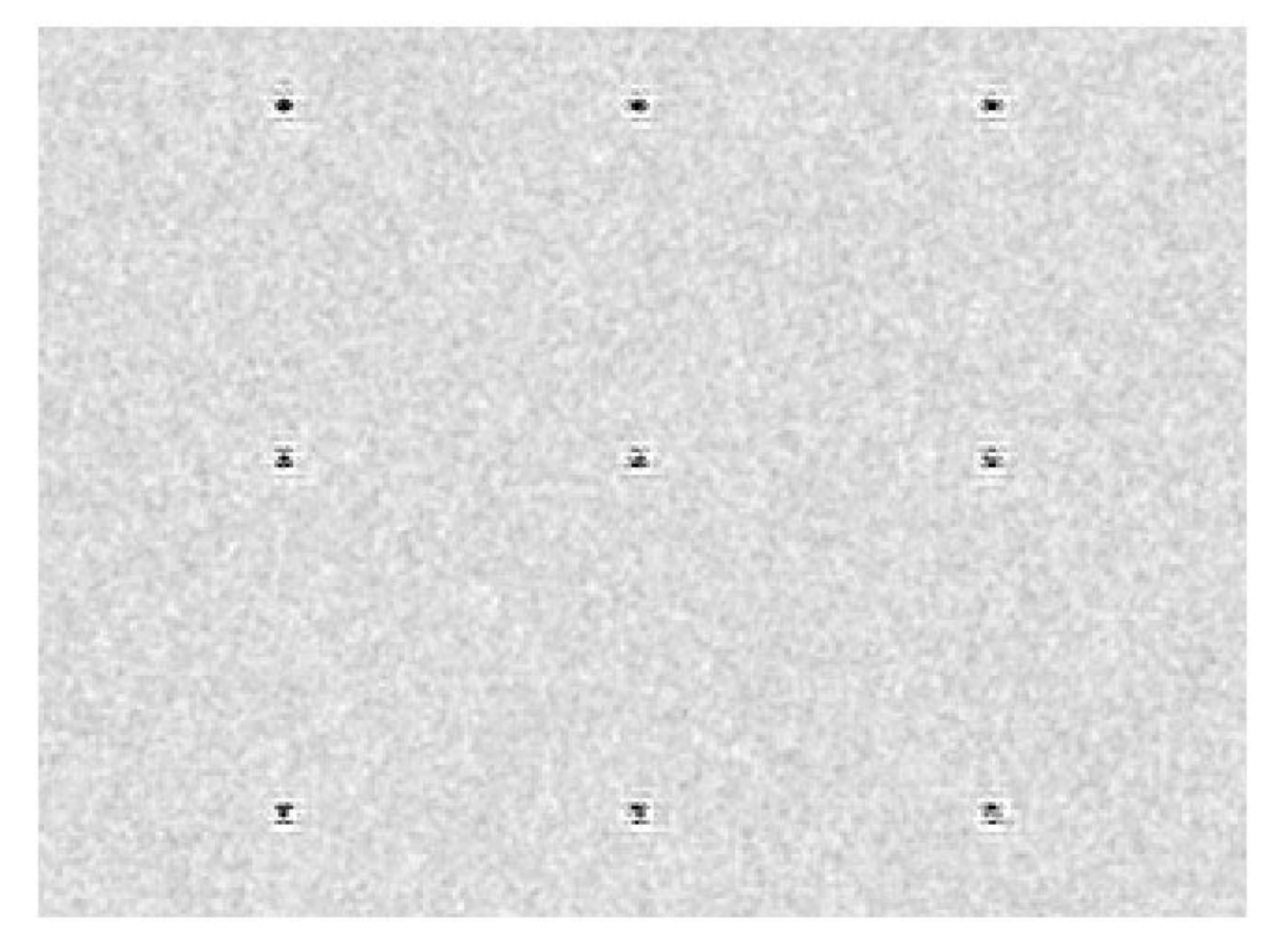}
\hspace{0. in}
\includegraphics[width=3.15in]{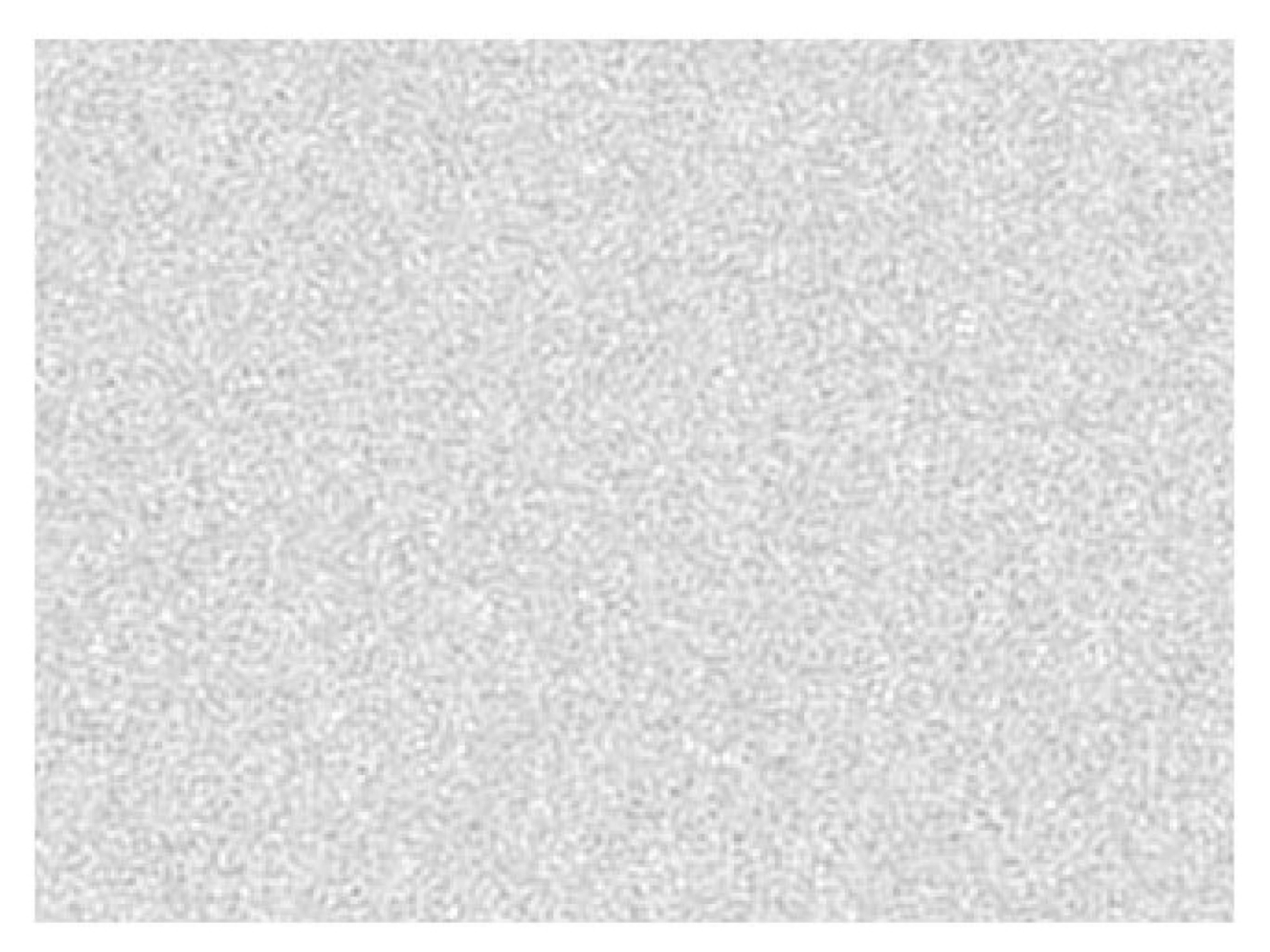}
\end{center}
%\plottwo{f6.eps}{f7.eps}
\caption{On the left is shown the difference image between a ``true'' WFPC2 image and the first approximation produced by the iDrizzle routine.   The simulated stars had a signal-to-noise ratio of about 100.  Twelve random pointing were used.   On the right is the difference of the iDrizzle image and the true image after two more iterations.    The stars have been completed removed.  No artifacts at the locations of stars are seen.  The subtraction is essentially limited by the statistical noise}
\label{fig-wfpc2-iter}
\end{figure}

In Figure~\ref{fig-ACS-iter} I show this procedure applied to simulated, noiseless ACS images.   A comparison with Figure~\ref{fig-ACS-driz} shows the large reduction of the drizzling artifacts in even the first full iteration, and the rapid convergence after that.   The reason for replacing the Voronoi approximation with Drizzle, however, was to handle cases with noise,  and therefore in Figure~\ref{fig-wfpc2-iter} I show the same result with simulated noisy WFPC2 images.   Again after a few iterations, all evidence of the stars is gone, and one sees only the noise of the sky.    

Even though the WFPC2 is heavily undersampled, the moderate signal-to-noise ratio for the original stars used in Figure~\ref{fig-wfpc2-iter} does not push the method to the limit of possible observations.  This is done more closely in the left-hand side of  Figure~\ref{UDF}, where I show the same subtraction performed on a combination of twelve simulated ACS images of the stellar field, with each star near saturation in a 1200s exposure, with appropriate Poisson and read noise added.   This tests the method in a situation of extremely high signal-to-noise, in essence the highest signal-to-noise ratio it is likely to face in practice.  The residuals are close to that expected from noise statistics, and the peak errors are reduced from the Drizzled subtraction by a factor of $\sim 20$.   However, the introduction of noise greatly slows convergence -- twenty-four iterations were used to produced the output shown here.  

On the right-hand-side of Figure~3 is a central region of the Hubble Ultra-Deep Field (Beckwith et al. 2003).   The bright star in this image is near saturation in each of the twelve 1200s individual exposures combined with the new method to form the final image.   When this image is mapped back onto the individual input exposures and subtracted, residuals of approximately 2\% peak are found under the stellar image.  These small but measureable errors are most likely caused by temporal variations in the PSF produced by the change in the insolation of the telescope as it orbits the earth.    Here, the residuals on the bright star were not improved by going beyond a few iterations.   

\section{Image Properties}

\subsection{Image Fidelity}

\begin{figure}[!h]
\begin{center}
\includegraphics[width=3in]{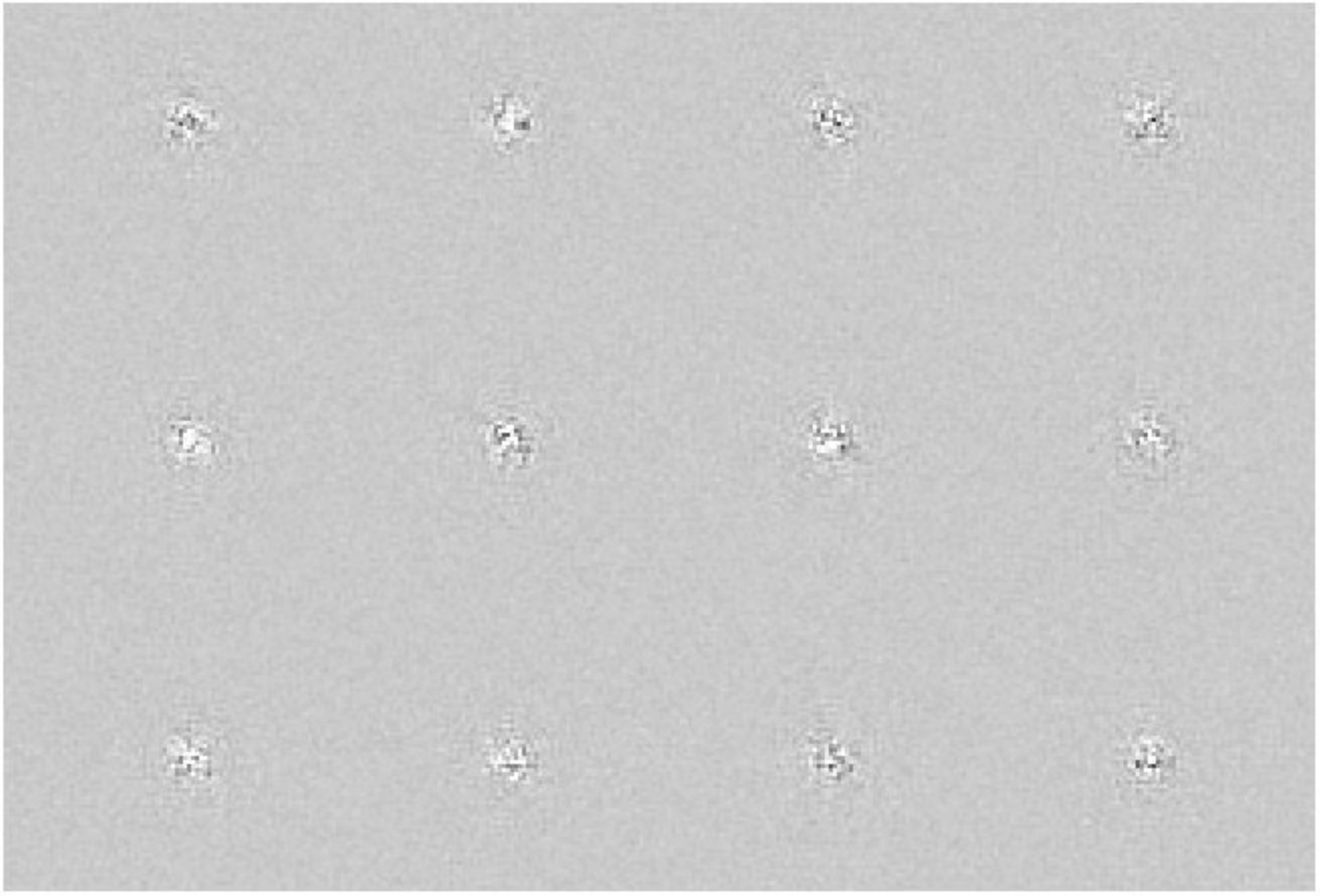}
\hspace{0.2 in}
\includegraphics[width=3in]{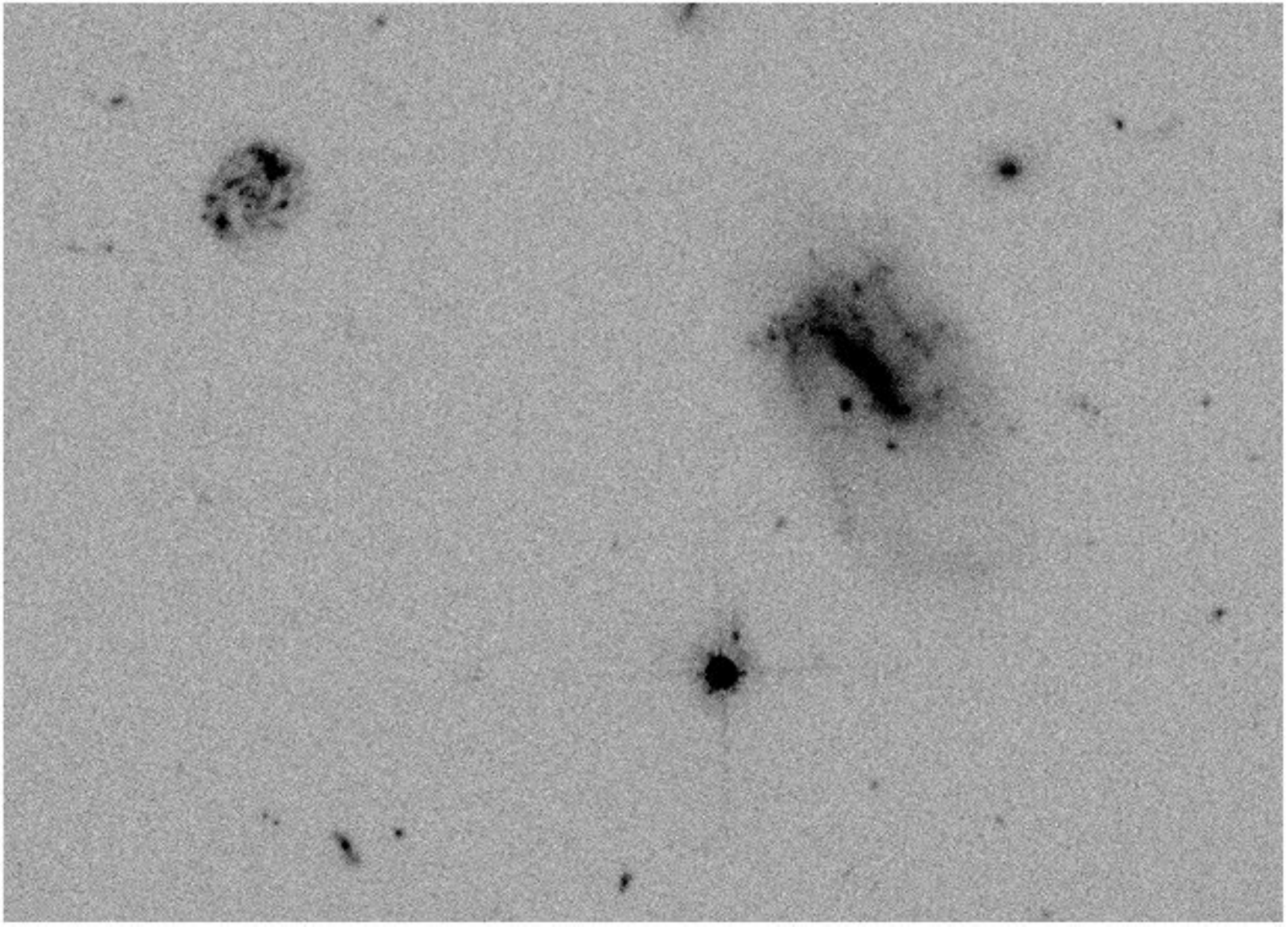}
\end{center}
%\plottwo{f8.eps}{f9.eps}
\caption{On the left, the same subtraction of the combination of twelve simulated ACS images from the true PSF as seen in Figures~\ref{fig-ACS-driz} and \ref{fig-ACS-iter} but with Poisson and read noise incorporated.  Here, the stars are far brighter than in Figure~\ref{fig-wfpc2-iter}, which shows the subtraction of noisy WFPC2 stellar images.  The stars are set just below image saturation in a single 1200s exposure of the F606W filter of the ACS.      Thus the Poisson noise of the stars overwhelms the sky and read noise of the images.  On the right, the combination of a section of twelve 1200s exposures in F606W from the Hubble Ultra-Deep Field using iDrizzle.   The bright star in the UDF image, just like those used in the simulation, is just below the ACS saturation limit.}
\label{UDF}
\end{figure}

As noted in the discussion of the method, iDrizzle will converge exactly in the absence of noise.  In the presence of noise, however,  successive iterations eventually lose their ability to improve the image as the power in the statistical noise overwhelms any systematic errors caused by the failure to converge to the correct underlying band-limited function.
For example, the noise in the combined image tested in left-hand side of Figure~\ref{UDF} is dominated by the Poisson noise of the bright stars in the original exposures.    This statistical noise limits the convergence of the method, and prevents the complete removal of  the faint ``ringing'' seen surrounding the centers of the (subtracted) point sources .  This is because two pixels from two different images may lie very close to one another on the sky, but have very different noise values.  Thus, the noise breaks the assumption that the data is truly band-limited, and the effect of this is most apparent in the Poisson noise of bright point sources.   However, the ringing in the image is  only $\sim 1 \times 10^{-4}$ of peak and is limited to a radius a few times the full-width at half-maximum (FWHM) of the PSF.   Without the subtraction of the original point source, the ringing is not visible. 

The test shown on the left-hand side of  Figure~\ref{UDF} was conducted using twelve simulated ACS images from HST with random sub-pixel dithers.   Good subtractions were seen with the ACS PSF and sampling with as few as eight random positions.   When the same test was repeated with simulated WFPC2 images, which are undersampled to a far greater degree, a minimum of twelve random positions was required before one saw good convergence.  At the same time, eight {\it optimally} placed dithers gave excellent results using WFPC2 sampling.  The ability of iDrizzle to accurately reproduce the PSF will depend on the ratio of the FWHM of the PSF to the size of the detector pixels, the number of samples that are available, and the pattern of those samples.  Additionally, because the removal of the systematic or pattern errors can be limited by the statistical noise power, deeper integrations will not only improve the statistical errors, they will also generally improve the systematic fidelity of the images.

\subsection{Statistical Noise Amplification}  

Drizzle places the center of a pixel output exactly where it was observed.    However, the average weight of an output pixel will not necessarily fall at the center of the pixel, and thus there is a jitter between the represented and effective position of a pixel.   Furthermore the peak of a drizzled PSF will never be greater than the greatest value in the appropriate region of the input images.  By contrast iDrizzle attempts to predict the true value of the image at the center of the output pixel, and thus the peak of a PSF will sometimes be brighter than any value of the input images in the appropriate region.   iDrizzle essentially uses estimates of the derivatives of the data to extrapolate to a position (the center of the output pixel) which is not necessarily exactly sampled in any of the input images.   This will produce some noise amplification, which will vary with the quality of the dithering.   In typical tests performed on this method the noise amplification has been in the region of 10\%.     This noise amplification, however, does not prevent the method from obtaining extraordinary results in high-signal to noise images.  For instance, when the PSF fitting software DAOPHOT (Stetson 1987)  is applied to the PSF based on the UDF bright star, it finds statistically identical photometry  when measuring the ``true'' representation of ACS PSFs (Figure 1) with purely statistical noise introduced and when measuring the iDrizzle reconstruction.   Both are at millimag accuracy and challenge the precision of the photometry software.   As the signal-to-noise level is lowered by using fainter simulated stars against a constant noise background, both the introduced noise and the noise surplus of about 10\% caused by the method become evident.  Nonetheless as can be seen in Figure 3 the image reconstruction remains very accurate.   The noise is dominated by statistical, rather than systematic, errors.

\subsection{ Correlated Noise and the Choice of the Inversion Filter}

Images produced by Drizzle show correlated noise.   Part of this correlation is caused by the drizzling process itself -- a non-zero value of  {\tt pixfrac} causes Drizzle to place a given input pixel value down on a region of linear size $p \times l$, where $p$ is the value of {\tt pixfrac} and $l$ is the linear size of an input pixel.  As the iteration proceeds, iDrizzle effectively forces $p$ to zero.    However, as noted earlier, noise correlation is introduced to the image by the application of the tapering function $L(\vec{k})$.    White noise is turned into reddish (correlated) noise by this procedure.   

Thus there is a competition between suppressing ringing near bright unresolved sources while maintaining optimal noise properties across the field.   This ringing is not caused by any power in the band-limited image passed by the telescope, but rather by the Poisson noise of the bright sources.   The limiting angular frequency of this power is not determined by the Nyquist limit of the original images, but rather by the spacing of the dithers of the combined images.  A gentle taper in Fourier space will best suppress this ringing.    However, to pass all of the information allowed by the optics, while limiting noise correlation, one prefers a sharp cuttoff near  the Nyquist frequency of $| \vec{k} | \approx \lambda/D$.   For the examples shown here a circular mask (in Fourier space) that falls from 90\% to 10\% transmission over a width of $0.1 \lambda/D$ is used.  The sampling
of the Drizzled image was nearly twice the Nyquist sampling of the band-limited image. This was found to give good suppression of ringing in the simulated WFPC2 and ACS images, while limiting correlated noise and minimizing the loss of true information in the image. The center of the taper function was slightly to the blue of the Nyquist frequency.   This essentially eliminates correlated noise (when a final image is created with Nyquist sampling), at the expense of allowing a slightly greater amount of noise through the filter.   No claim is made that this particular choice of cutoff, apodization, or sampling is optimal, but it was found effective.   

\subsection{Determining Convergence}

The algorithm described here is part of a class of techniques that vary in their use of specific
weightings or interpolation schemes to iteratively arrive at a solution of the irregular sampling 
problem.  These all tend to converge geometrically to the correct solution with the speed
of conversion being dependent on the value of $2 l \nu_0$, where $l$ is the maximal distance
between samples and $\nu_0$ is the Nyquist frequency of the data.   Where  $2 l \nu_0 \ll 1$, 
convergence is rapid; however, where $2 l \nu_0 \sim 1$, convergence is slow
or fails (see Werther 1999, Eq 50 for a specific case, and
Werther 2004 and Feichtinger. \& Gr\"ochening 1994 for broader discussions).     

In the case being discussed here, however, there is no unique solution because of the presence
of noise.   Further, because of geometric distortion the value of  $l$, and thus $2 l \nu_0$,  is 
likely to vary across the image, and thus so too will the rapidity of local convergence.   

In practice, it is fairly simple for the user to test whether a particular observing plan with a given
instrument is likely to produce acceptable images.  To do this  the user should create a well sampled ``output'' image.  In the test cases shown here, PSFs placed at regular intervals with a small random dither were used.   The user can vary the amplitude of the PSF as well, to see how the technique works in different noise regimes.  This simulated image is then mapped to the input image scale using Drizzle package routine, {\tt blot}.   {\tt blot} will, with the correct distortion files, map the image back as it would have appeared on the distorted field of the detector.   The user can create as many input images as he or she desires, with a dither pattern emulating the one expected to be obtained from the observations.   Noise can then be added to each image to match the level expected in the observations.   Finally one can combine the images using iDrizzle.   This will allow the user to check how many iterations it will take to obtain a reasonable result.    As the amplification of the noise tends to grow with the number of iterations, greatly exceeding the number of required iterations is not advised.   Usually a few to a dozen interations will do.   Indeed, in the case of the real UDF data, convergence stopped after only a few iterations.  As mentioned earlier, this is because the technique was limited by the variability of the HST PSF.

\section{Discussion and Future Developments}

In this paper, a new method for the combination of dithered astronomical images has been introduced.   The method, iDrizzle, has the ability to handle shifts, distortions and missing data, and converges rapidly to an accurate representation of the underlying image.  It provides a dramatic improvement in fidelity over Drizzle, particularly on resolved objects, at the cost of a small increase in statistical noise.   It requires, however, that the combined images come close to Nyquist sampling across  the combined image. 

The algorithm upon which this worked is in part based, Drizzle, by itself creates excellent images of {\it resolved} objects (such as galaxies in a deep HST image), and will produce images of slightly higher signal-to-noise images than iDrizzle, in the case that statistical noise is larger than the small artifacts of using Drizzle.   Similarly, observers attempting precise stellar photometry, may use Drizzle and/or iDrizzle to locate their stars and remove image defects,  and can do photometry on each individual image and combine results to avoid the small increase in statistical noise produced by iDrizzle.  However, this approach would require a substantial amount of effort to avoid the inherent photometric biases of working with undersampled data -- something that iDrizzle does directly.  Where iDrizzle truly stands out is in creating accurate images of objects with unresolved or nearly-unresolved components.  As the test with the UDF data showed, iDrizzle allows far more accurate image subtraction, and thus may be of great use to observers trying to remove powerful point sources underlying extended objects, such as AGN from their host galaxies.   In the long-run though, an important use of iDrizzle may be the most direct application, the creation of point-source functions.   

Future space-based wide-field imagers are likely to be undersampled.   Nonetheless they will be used (and some will be expressly intended) for applications such as lensing which require excellent knowledge of the PSF.   As shown here, with a moderate number of dithers, even with random sampling, iDrizzle can produce high-fidelity PSFs.    The desire to cover a large area of sky, however, may mean that  some projects will prefer to keep the number of dithers in a given filter to a bare minimum -- four exposures may be a common preference.   While four dithers is too few for iDrizzle to estimate the image in the absence of near regular sampling,  iDrizzle could be used to recover the PSF if the telescope has either temporal or spatial stability.  With temporal stability, images of different stars with similar colors taken near the same location on the detector could be combined to form a single PSF.   Similarly, if the PSF is stable over a sufficiently wide field, different stars of similar color from a single image may be combined as a substitute for an increased number of dithers.   While in both cases one will have to fit for the magnitudes and relative positions of the stars, as magnitudes and positions are identical (up to a known shift) for all dither sets of the same star, the small number of parameters fits should typically add little noise, particularly if the fitting is done iteratively while solving for the shape of the PSF.

The new method presented here, iDrizzle, may find use in a number of applications where high image fidelity is required.  In order that the community may test this algorithm and expand upon its use, the Pyraf script used to create the UDF image, along with the UDF data used in this test, is posted on the web for download by interested readers at http://www.stsci.edu/\textasciitilde fruchter/iDrizzle/UDF-iDrizzle .

\section{Acknowledgements}

I wish to thank Robert Lupton for helpful and amusing discussions, the Institute of Pure and Applied Mathematics at UCLA for hosting a workshop on astronomical imaging where I was first able to discuss some of these ideas, and the Space Telescope Science Institute, which supported this work.  Hans-Martin Adorf and Thomas Strohmer gave very useful thoughts on analytical methods related to irregular sampling.  Anton Koekemoer provided the aligned UDF data, and helped with the first Python implementation of the iDrizzle algorithm.   I also thank the referee, Tod Lauer, as well as Stefano Casertano, who both provided comments that lead to a clearer final manuscript.

%\clearpage
%\break

\clearpage

\clearpage

\clearpage

\clearpage

\begin{thebibliography}{}
\bibitem[Ald(2001)]{Ald} Aldroubi, A. \& Gr\"ochening, K. 2001, SIAM Rev., 43, 585
\bibitem[Beck(1982)]{Beck} Beckwith, S. V. W. et al. 2006, \aj, 684, 1404
\bibitem[Bene(1991)]{Bene}Benedetto, J.J. 1992,  in Wavelets: A Tutorial in Theory and Applications (San Diego: Academic Press), 445
\bibitem[Fecht(1994)]{Fecht}Feichtinger, H. G. \& Gr\"ochening, K. 1994, in Wavelets: Mathematics and Applications, Benedetto, J. \& Frazier, M, eds. (Boca Raton: CRC Press), 305
\bibitem[Fruc(2002)]{Fruc} Fruchter, A. S. \& Hook, R. N. 2002, \pasp, 114, 144
\bibitem[GaS(2001)]{GaS}Gr\"ochenig, K. \& Strohmer, T. 2001, in Nonuniform Sampling: Theory and Practice (New York: Kluwer)
\bibitem[Kotel(1933)]{Kotel} Kotelnikov, V.A., 1933, in Proceedings of the First All-Union Conference on the Technological Reconstruction of the Communications Sector and the Development of Low-current Engineering (Moscow: Izd. Redl. Upr. Svyanki RKKA), http://ict.open.ac.uk/classics/1.pdf
\bibitem[Lauer(1999)]{Lauer} Lauer, T. R. 1999, \pasp, 111, 227
\bibitem[Shan(1948)]{Shan} Shannon, C.E., 1948, Bell System Technical J., 27, 379
\bibitem[Stet(1987)]{Stet} Stetson, P. 1987, \pasp, 99, 191
\bibitem[Wert(1999)]{Wert} Werther, T. 1999, {\it Reconstruction from Irregular Samples with Inproved Locality}, http://citeseerx.ist.psu.edu/viewdoc/summary?doi=10.1.1.28.402
\bibitem[Wert(2004)]{Wert2} Werther, T. 2004, {\it A First Guided Tour on the Irregular Sampling Problem}, http://www.math.ucdavis.edu/~strohmer/research/sampling/irsampl.html
\bibitem[Whit(1935)]{Whit} Whittaker, J.M., 1935, {\it Interpolatory Function Theory}, Cambridge Univ. Press
\bibitem[Will(1996)]{Will} Williams, R. E. et al. 1996, \aj, 112, 1335
\end{thebibliography}
\end{document}